\documentclass[prd,twocolumn,nofootinbib,superscriptaddress]{revtex4-2}
\usepackage{soul}
\usepackage{preamble}
\usepackage{multirow} 
\newcommand{\mnras}{Mon. Not. Roy. Astr. Soc.}
\newcommand{\sovast}{Soviet Astronomy}
\newcommand{\ssr}{Space Science Reviews}
\newcommand{\aap}{Astronomy and Astrophysics}

\newcommand{\apjl}{Astrophysical Journal, Letters}
\newcommand{\apjs}{Astrophysical Journal, Supplement}

\begin{document}
\title{
Signature of Particle Diffusion on the X-ray Spectra of the blazar Mkn 421
}
\author{C. Baheeja}
\email{baheeja314@gmail.com}
\affiliation{Department of Physics, University of Calicut, Malappuram, Kerala, India}
\author{S. Sahayanathan}
\email{sunder@barc.gov.in}
\affiliation{Astrophysical Sciences Division, Bhabha Atomic Research Centre, Mumbai - 400085, India}
\affiliation{Homi Bhabha National Institute, Mumbai 400094, India}
\author{F.M. Rieger}
\email{frank.rieger@ipp.mpg.de}
\affiliation{Max Planck Institute for Plasma Physics (IPP), Boltzmannstra{\ss}e 2,
85748 Garching, Germany}
\affiliation{Institute for Theoretical Physics, Heidelberg University, Philosophenweg 12, 
69120 Heidelberg, Germany}
\author{C.D. Ravikumar}
\email{cdr@uoc.ac.in}
\affiliation{Department of Physics, University of Calicut, Malappuram, Kerala, India}
\begin{abstract}

The  curvature in  blazar spectrum has the potential to understand the particle dynamics in jets. We performed a detailed analysis of simultaneous \emph{Swift}-XRT (0.3-10 keV) and \mbox{\emph{NuSTAR} (3-79 keV)} observations of Mkn 421. Our analysis of \emph{NuSTAR} observations alone reveals that, during periods of low flux, the hard X-ray spectra are best represented by a steep power-law with photon index reaching $\sim$ 3. However, the spectrum exhibits significant curvature during its high flux states. To investigate this, we explore plausible diffusion processes facilitating shock acceleration in the emission region that can contribute to the observed spectral curvature. Particularly, such processes can cause gradual fall of the photon spectrum at high energies which can be represented by a sub-exponential function. The parameter that decides this spectral change can be used to characterise the energy dependence of the diffusive process. Our results suggest that the X-ray spectra of Mkn 421 are consistent with a scenario where particle acceleration is mediated through Bohm-type diffusion and the spectra beyond the synchrotron peak is modulated by the radiative loss process.

\end{abstract}
\maketitle
\section{Introduction}
\label{sec_intro}
Blazars are a class of Active Galactic Nuclei (AGNs) with a relativistic jet closely
aligned to the observer’s line of sight \citep{doi:10.1146/annurev.aa.31.090193.002353,Urry_1995}. They exhibit rapid flux variations at different time-scales, spanning from minutes to years, across the entire electromagnetic spectrum \citep{doi:10.1146/annurev.astro.35.1.445}. These highly variable sources are characterized by two prominent peaks in their spectral energy distribution (SED), extending from radio to $\gamma-$ray energies. The low-energy SED component (peaking at the IR to X-ray regime) is commonly attributed to synchrotron emission by relativistic electrons \citep{10.1093/mnras/241.1.43P,Urry_1995}. On the other hand, the physical process responsible for the emission of the high-energy SED component (peaking at the MeV to TeV gamma-ray regime) is still under debate \citep[e.g.,][for a review]{Sol2022}. 

Modeling the broadband synchrotron spectral component of blazars often suggests the 
emitting electron distribution to be a broken power-law \citep{1994ApJS...95..371S,
Tavecchio_1998,Fossati_2000}. In principle, a power-law electron distribution could be 
produced by Fermi-type particle acceleration \citep[e.g.,][]{Fermi1949,Lemoine2019}, 
where particles gain energy as they are scattered by magnetic turbulence structures 
embedded in the jet. When electrons are scattered by the turbulent structures across 
a shock front, Fermi acceleration will be efficient and is commonly referred to as 
shock acceleration \citep[e.g.,][]{1998A&A...333..452K,Rieger_2007}. Synchrotron 
cooling of a power-law electron distribution then results in a broken power-law 
distribution with power-law indices differing by one. The corresponding synchrotron 
photon spectrum would follow a broken power-law with indices differing by half 
\citep{1962SvA.....6..317K,1986rpa..book.....R}. However, broadband SED modeling of 
blazars often fails to support such an interpretation \citep[e.g.,][]{Mankuzhiyil_2012}. 
In particular, our earlier study of the blazar Mkn 421 reported a strong 
anti-correlation between the spectral indices measured at lower and higher energies 
around the synchrotron spectral peak, disfavouring a simple radiative loss origin 
\citep{2022MNRAS.514.3074B}. 

Narrow-band spectral analysis of blazars frequently reveals significant curvature 
around the peak of the synchrotron component, that is formally well represented by 
a log-parabola function \citep{1986ApJ...308...78L,2004A&A...413..489M,
2007A&A...467..501T,Chen_2014}. However, such a function fails to explain the 
combined optical-UV and X-ray spectrum \citep{2004A&A...413..489M,2009A&A...501..879T,
2015A&A...580A.100S,2018MNRAS.478L.105J}. Instead, a smooth broken power-law or a 
power-law particle distribution with an exponential cutoff is often capable of 
explaining this broad-band spectral component \citep{Sinha_2017,1998A&A...333..452K}. 
Recent studies with high-resolution observations have revealed significant spectral 
curvature even beyond the synchrotron peak \citep{10.1093/mnras/stu1703,
2015A&A...580A.100S,Gaur_2017}.

A simple description of first-order Fermi acceleration at shocks, assuming an 
energy-independent acceleration and escape time scale, naturally produces a 
power-law electron distribution. However, when radiative losses are taken into 
account and an energy-dependence is incorporated into the acceleration and/or 
escape time scales, the resulting particle distribution can deviate from a power-law 
and exhibit curvature towards high energies \citep{1998A&A...333..452K}. Similarly, 
a log-parabolic photon spectrum suggests the underlying electron distribution to 
be log-parabolic, which could be interpreted in terms of a statistical acceleration 
scenario with an energy-dependent acceleration probability \citep{2004A&A...413..489M}. 
In fact, under specific conditions, the electron distribution resulting from a
stochastic particle acceleration could mimic a log-parabola  \citep{2011ApJ...739...66T}. 
It is then expected that the synchrotron peak energy anti-correlates with spectral curvature 
\citep{2007A&A...466..521T}. During the period of 1997$-$2006, an anti-correlation between 
the synchrotron peak energy and the spectral curvature was indeed found through log-parabolic spectral fitting of the X-ray data from Mkn 421 and other TeV BL Lacs, seemingly supportive of a stochastic acceleration scenario \citep{2007A&A...466..521T,2009A&A...501..879T,2008A&A...478..395M,2011ApJ...739...66T}. However, recent studies using \emph{Swift}$-$XRT/\emph{NuSTAR} observations report no significant correlation between these quantities \citep{2018ApJ...854...66K,2018ApJ...858...68K,Kapanadze_2020,2015A&A...580A.100S,2022MNRAS.514.3074B}. On the other hand, a curved spectrum could also be the outcome of an energy-dependent escape from the acceleration region. When this energy-dependence is mild, the resulting electron distribution closely follows a log-parabolic shape \citep{2018MNRAS.478L.105J} but deviates significantly otherwise. 
Synchrotron emission by an electron distribution originating in a model with a strong energy-dependent escape time scale has been used to fit the spectra of Mkn~421 during different flux states \citep{10.1093/mnras/sty2003,Goswami_2020}. A strong correlation was observed between flux and the  energy-dependence of the escape time-scale, and this supports 
that Bohm type diffusion is prominent during high flux states.

Particle acceleration at non- or mildly relativistic shock fronts has for long been 
considered as one of the preferred mechanisms for generating the non-thermal particle 
distributions seen in AGN jets \citep[e.g.,][]{Marscher1985,1998A&A...333..452K,
Rieger_2007,Zech2021,DiGesu2022,DiGesu2023}. 
The highest energy achieved by the accelerated particles, as well as the shape of the 
spectrum around the maximum energy, are influenced by the balance between acceleration, 
escape and the radiative energy loss rates. In the presence of synchrotron losses, shock acceleration (for example) can result in a power-law particle distribution with a 
modified exponential cutoff, $\propto \exp[-(\gamma/\gamma_c)^{\beta_e}]$, where 
$\beta_e$ is dependent on the underlying turbulence/diffusion properties \citep[e.g.,][]
{Zirakashvili_2007}. Formally, $\beta_e =(1+a)$ is related to the momentum index $a$ of 
the spatial diffusion coefficient, $\kappa=(1/3) \lambda\,c \propto \gamma^a$, that 
facilitates the particle transport. Here, $\lambda$ is the particle mean free path 
and $\gamma$ is the particle Lorentz factor. In particular, one may have $\beta_e = 
1$ ($a = 0$) in the case of \mbox{“idealized”} hard-sphere scattering (energy-independent 
diffusion), $\beta_e = 4/3$ ($a = 1/3$) for Kolmogorov-type turbulence, and $\beta_e 
= 2$ ($a = 1$) for Bohm type diffusion (where $\lambda \sim r_g$, with $r_g$ as 
the gyro-radius). Since the corresponding particle acceleration timescale, $t_{\rm acc}$ is proportional to $\lambda $, Bohm diffusion typically yields the fastest acceleration 
rate (i.e., the highest $\gamma_{\rm e,max}$ when balanced with synchrotron losses). The 
resultant synchrotron spectra can exhibit some extended curvature at high energies. 
In particular, an electron distribution with exponential cutoff index $\beta_e$
will result in a synchrotron spectrum which can be significantly smoother 
(sub-exponential) $j_\nu \propto \rm exp\left[-(\nu/\nu_c)^{\zeta}\right]$ with $\zeta 
\equiv \frac{\beta_e}{\beta_e+2}$, e.g., $\zeta = 1/2$ in the case of Bohm-type diffusion
\citep{1989A&A...214...14F}.

The BL Lac object Mkn~421, that we focus on here, is the nearest (z = 0.031) and one 
of the well-studied TeV blazars. Mkn~421 belongs to the high-frequency BL Lac (HBL) 
class, as its synchrotron spectral component peaks in the X-ray regime. The X-ray 
spectrum around the synchrotron peak exhibits significant curvature that has been
interpreted in terms of a log-parabola function. X-ray spectral analysis of Mkn~421 
using \emph{NuSTAR} (3-79 keV) observations, reveals that during the low-flux state of the 
source in January 2013, the hard X-ray spectra were well represented by a steep 
power-law model with a photon index saturating at $\sim 3$  \citep{Kataoka_2016,
Balokovi__2016}. However, the observed X-ray spectrum also shows a significant 
curvature during high-flux states in April 2013 \citep{2015A&A...580A.100S}. This 
curvature persists even in the hard X-rays, which makes it (in spite of the fact
that some blending of components cannot be excluded) challenging to attribute this 
solely to the spectral transition occurring at the peak of the synchrotron component 
\citep{Fossati_2008,Horan_2009}. 
To explore this further, we have performed a detailed spectral study on the 
X-ray data of the source. We are particularly interested to understand whether the 
observed X-ray spectral characteristics allow some inferences on the turbulence 
properties in the jet. 
The paper is organized as follows: In Section \S \ref{sec_obs}, we discuss the 
observation and data reduction procedure, while the X-ray spectral study is 
described in Section \S \ref{sec_spectral-fit}. The summary is presented in 
Section \S \ref{sec_discussion}.

\section{Observation and Data Analysis}
\label{sec_obs}
Mkn 421 has been observed by \emph{NuSTAR} and \emph{Swift}-XRT in both flaring as well as quiescent 
flux states. For the current study we have selected all the available simultaneous 
\emph{Swift}-XRT and \emph{NuSTAR} observations till 2018 (details are given in Table 
\ref{tab:NuSTAR-simul}). This allows us to analyse the source over a wide range of X-ray 
energies, from 0.3 to 79 keV. The strategies for analysing these observations are detailed 
below.
\begin{table*}
\begin{center}
\caption{ Details of simultaneous \emph{Swift}-XRT and \emph{NuSTAR} observations.} 
\begin{tabular}{ c c c c c c c}        
\hline
\emph{Swift}	&	Date \& time	&	Exposure			&	\emph{NuSTAR}	&	Date \& time	&	Exposure	\\
 Obs.ID	&		&	(s)			&	Obs.ID	&		&	(s)	\\
\hline 
 
35014034	&	2013-01-15 02:09:59	&	3958.859	&	60002023006	&	2013-01-15 00:56:07	&	24181	\\
80050003	&	2013-02-06 01:20:59	&	9506.827	&	60002023010	&	2013-02-06 00:16:07	&	19302	\\
80050006	&	2013-02-17 00:03:59	&	9201.642	&	60002023014	&	2013-02-16 23:36:07	&	17356	\\
80050007	&	2013-03-04 23:34:25	&	984.609	&	60002023016	&	2013-03-04 23:06:07	&	17251	\\
80050011	&	2013-03-11 23:58:59	&	8425.937	&	60002023018	&	2013-03-11 23:01:07	&	17472	\\
80050013	&	2013-03-17 01:22:59	&	8880.74	&	60002023020	&	2013-03-17 00:11:07	&	16554	\\
80050014	&	2013-04-02 21:01:59	&	1644.569	&	60002023022	&	2013-04-02 17:16:07	&	24767\\	
80050016	&	2013-04-11 00:30:59	&	1118.631	&	60002023024	&	2013-04-10 21:26:07	&	5757	\\
80050019	&	2013-04-12 21:53:58	&	9546.279	&	60002023027	&	2013-04-12 20:36:07	&	7629	\\
32792002	&	2013-04-14 00:38:59	&	6327.071	&	60002023029	&	2013-04-13 21:36:07	&	16508	\\
35014062	&	2013-04-15 23:07:59	&	534.621	&	60002023033	&	2013-04-15 22:01:07	&	17276	\\
35014065	&	2013-04-17 00:46:59	&	8842.132	&	60002023035	&	2013-04-16 22:21:07	&	20278	\\
35014066	&	2013-04-18 00:49:59	&	6887.219	&	60002023037	&	2013-04-18 00:16:07	&	17795	\\
35014067	&	2013-04-19 00:52:59	&	6132.768	&	60002023039	&	2013-04-19 00:31:07	&	15958	\\
34228110	&	2017-01-04 00:06:57	&	6021.027	&	60202048002	&	2017-01-03 23:51:09	&	23691	\\
81926001	&	2017-01-31 23:27:57	&	1009.619	&	60202048004	&	2017-01-31 23:46:09	&	21564	\\
34228145	&	2017-02-28 22:46:56	&	44.62	&	60202048006	&	2017-02-28 22:11:09	&	23906	\\
\hline
\end{tabular}
\label{tab:NuSTAR-simul} 
\end{center} 
\end{table*}
\subsection{\emph{NuSTAR}}
\emph{NuSTAR} \citep{Harrison_2013} is a space-based hard X-ray telescope which operates from 3 
to 79 keV energy band with an angular resolution of subarcmin. All observations are carried 
out with two co-aligned, independent telescopes called  Focal Plane Module A (FPMA)
and B  (FPMB). The \emph{NuSTAR} observations were taken from the HEASARC interface by NASA and the data were processed with  NuSTARDAS package (Version 1.4.1) available within HEASOFT 
(Version 6.19). The source spectrum is extracted from a circular region with a radius of 
50 arcsec centered on the source, while the background is estimated from a circular region 
with a radius of 70 arcsec that is free of source contamination but near it. NUPRODUCT 
(Version 0.2.8) was used to obtain source and background spectra after running NUPIPELINE 
(Version 0.4.9) on each observation. The FPMA and FPMB source spectra were then individually 
grouped to 30 photons per bin using the 
tool GRPPHA to ensure improved $\chi^2$ statistics. 
\subsection{\emph{Swift}-XRT}
The XRT is a focusing X-ray telescope operating in the 0.3-10 keV energy range with an angular resolution $<20$ arcsec \citep{2005SSRv..120..165B}. The \emph{Swift}-XRT observations were also retrieved from HEASARC interface and the data processed using the XRTDAS software package (Version 3.0.0) available within HEASOFT. We used the observations performed in Windowed Timing (WT) mode and the events with 0-2 grades have been considered in the analysis. The event files were cleaned and calibrated using standard procedures with the XRTPIPELINE (Version v0.13.4) task. A circular region of 30 pixel radius centred at the source was used to extract the source spectrum, and a circular region of same size devoid of source contamination was used to extract the background spectrum. An annular region with inner and outer radii of 2  and 30 pixels, respectively,  was used as source and background regions for the observation with  pileup (Obs.ID 80050019). XRTPRODUCTS (Version v0.4.2) was used to generate the final spectrum. The XRTMKARF task was employed to generate the auxiliary response files (ARFs), and the response matrice files (RMFs) from the \emph{Swift} CALDB were used.  The source spectra were then 
grouped using the GRPPHA tool to ensure a minimum of 20 counts/bin.

\section{X--ray spectral analysis and results}\label{sec_spectral-fit}
\subsection{\emph{NuSTAR} (3-79 keV) regime}
To investigate the curvature in the hard X-ray regime, we fitted the \emph{NuSTAR} observations (3-79 keV) 
of the source, using three models available in XSPEC namely, power-law (PL), log-parabola (LP), 
and a power-law with an exponential cutoff (CPL). These models are defined as
\begin{align}
	F(\epsilon)\propto \epsilon^{-\Gamma} \quad\quad \mathrm{(PL)}\,,
\end{align}
where $\epsilon$ is the photon energy, and $\Gamma$ represents the power-law index
\begin{align}
	F(\epsilon)\propto 
     \left(\frac{\epsilon}{\epsilon_0}\right)^{-\alpha-\beta\,\rm{log}(\epsilon/\epsilon_0)} 
     \quad\quad \mathrm{(LP)}\,,
\end{align}
where $\alpha$ is the spectral slope at energy $\epsilon_0$, and $\beta$ is the spectral 
curvature, and 
\begin{equation}
 F(\epsilon) \propto \epsilon^{-p}\, \rm exp[-(\epsilon/\epsilon_c)]
  \quad\quad \mathrm{(CPL)}\,,
\end{equation}
where $p$ represents the power-law index, and $\epsilon_c$ characterizes the position of 
the cutoff energy. The spectral peak ($\epsilon^2 F(\epsilon)$ representation) of the 
log-parabola function is obtained from
\begin{equation}
\label{eq:ep}
 \epsilon_p = \epsilon_0\,10^{\frac{2-\alpha}{2\,\beta}}\,.
 \end{equation}
 \begin{table*}
\begin{center}
\caption{ Fit parameters of \emph{NuSTAR} (3-79 keV) spectra as modeled with PL, 
LP and CPL.} 
 \begin{tabular}{cc cc ccc cccccc} 
\hline
	\emph{NuSTAR}	&Flux	&&	\multicolumn{2}{c} {PL} 	&&	\multicolumn{3}{c} {LP($\epsilon_0$=5 keV)}	&&	\multicolumn{3}{c} {CPL}\\	
\cline{4-5}
 \cline{7-9}
 \cline{11-13}	 
		Obs.ID		&	(3-79 keV) &&	$\Gamma$ & $\chi_{red}^2$	&&	$\alpha$	&	$\beta$	& $\chi_{red}^2$ &&	$p$	&	$\epsilon_c$ (keV) & $\chi_{red}^2$ \\
\hline
\hline
60002023006	&	-9.954$\pm$0.004	&	&	3.03$\pm$0.01	&	1.05 (563)	&	&	2.95$\pm$0.02	&	0.31$\pm$0.06	&	0.92 (562)	&	&	2.82$\pm$0.05	&	35.12$^{+9.75}_{-6.47}$		&	0.94 (562)	\\
60002023010	&	-9.842$\pm$0.004	&	&	2.95$\pm$0.01	&	1.32 (570)	&	&	2.83$\pm$0.02	&	0.41$\pm$0.06	&	1.04 (569)	&	&	2.64$\pm$0.05	&	24.99$^{+4.49}_{-3.41}$		&	1.06 (569)	\\
60002023014	&	-10.187$\pm$0.007	&	&	3.02$\pm$0.02	&	1.06 (419)	&	&	3$\pm$0.03	&	0.11$\pm$0.09	&	1.05 (418)	&	&	$-$	&	$-$	&	$-$	\\		
60002023016	&	-9.779$\pm$0.004	&	&	3.01$\pm$0.01	&	1.11 (557)	&	&	2.95$\pm$0.02	&	0.25$\pm$0.06	&	1.01 (556)	&	&	2.85$\pm$0.05	&	44.97$^{+16.44}_{-9.8}$		&	1.03 (556)	\\
60002023018	&	-9.906$\pm$0.005	&	&	3.09$\pm$0.02	&	1.04 (509)	&	&	3.04$\pm$0.02	&	0.2$\pm$0.07	&	0.99 (508)	&	&	2.96$\pm$0.05	&	53.64$^{+31.18}_{-14.91}$		&	1 (508)	\\
60002023020	&	-9.731$\pm$0.005	&	&	2.77$\pm$0.01	&	1.18 (595)	&	&	2.68$\pm$0.02	&	0.28$\pm$0.05	&	1.04 (594)	&	&	2.58$\pm$0.04	&	40.77$^{+10.57}_{-7.19}$		&	1.06 (594)	\\
60002023022	&	-9.343$\pm$0.002	&	&	2.74$\pm$0.01	&	1.51 (898)	&	&	2.64$\pm$0.01	&	0.3$\pm$0.03	&	1.03 (897)	&	&	2.52$\pm$0.02	&	38.28$^{+3.89}_{-3.29}$		&	1.06 (897)	\\
60002023024	&	-9.201$\pm$0.003	&	&	2.9$\pm$0.01	&	1.47 (620)	&	&	2.78$\pm$0.02	&	0.39$\pm$0.05	&	1.11 (619)	&	&	2.61$\pm$0.04	&	27.88$^{+4.2}_{-3.33}$		&	1.14 (619)	\\
60002023027	&	-8.624$\pm$0.002	&	&	2.62$\pm$0.01	&	2.63 (1023)	&	&	2.45$\pm$0.01	&	0.44$\pm$0.02	&	0.94 (1022)	&	&	2.29$\pm$0.02	&	26.4$^{+1.36}_{-1.26}$		&	1.01 (1022)	\\
60002023029	&	-9.085$\pm$0.002	&	&	2.79$\pm$0.01	&	2.06 (917)	&	&	2.65$\pm$0.01	&	0.42$\pm$0.02	&	0.98 (916)	&	&	2.48$\pm$0.02	&	26.67$^{+1.83}_{-1.64}$		&	1.08 (916)	\\
60002023033	&	-9.014$\pm$0.002	&	&	2.59$\pm$0.01	&	1.86 (1019)	&	&	2.46$\pm$0.01	&	0.33$\pm$0.02	&	1.01 (1018)	&	&	2.34$\pm$0.02	&	35.89$^{+2.5}_{-2.23}$		&	1.02 (1018)	\\
60002023035	&	-8.934$\pm$0.002	&	&	2.39$\pm$0.01	&	2.33 (1182)	&	&	2.25$\pm$0.01	&	0.35$\pm$0.02	&	1.07 (1181)	&	&	2.13$\pm$0.01	&	35.77$^{+1.9}_{-1.74}$		&	1.1 (1181)	\\
60002023037	&	-9.835$\pm$0.004	&	&	2.85$\pm$0.01	&	1.31 (568)	&	&	2.72$\pm$0.02	&	0.4$\pm$0.06	&	1.05 (567)	&	&	2.55$\pm$0.05	&	26.61$^{+4.85}_{-3.67}$		&	1.05 (567)	\\
60002023039	&	-9.868$\pm$0.005	&	&	2.94$\pm$0.02	&	0.98 (519)	&	&	2.88$\pm$0.02	&	0.22$\pm$0.06	&	0.91 (518)	&	&	2.79$\pm$0.05	&	50.09$^{+23.58}_{-12.58}$		&	0.92 (518)	\\
60202048002	&	-9.269$\pm$0.003	&	&	2.45$\pm$0.01	&	1.39 (1006)	&	&	2.36$\pm$0.01	&	0.22$\pm$0.02	&	1.1 (1005)	&	&	2.29$\pm$0.02	&	60.22$^{+7.58}_{-6.17}$		&	1.14 (1005)	\\
60202048004	&	-9.247$\pm$0.003	&	&	2.45$\pm$0.01	&	1.72 (1001)	&	&	2.31$\pm$0.01	&	0.33$\pm$0.02	&	1.04 (1000)	&	&	2.2$\pm$0.02	&	37.22$^{+2.94}_{-2.59}$		&	1.06 (1000)	\\
60202048006	&	-9.261$\pm$0.002	&	&	2.49$\pm$0.01	&	1.75 (996)	&	&	2.37$\pm$0.01	&	0.31$\pm$0.02	&	1.13 (995)	&	&	2.26$\pm$0.02	&	39.53$^{+3.31}_{-2.89}$		&	1.16 (995)	\\
\hline
\end{tabular}   
\label{tab:nustar-alone-fit} 
\end{center}   
\end{table*}
\begin{figure*}
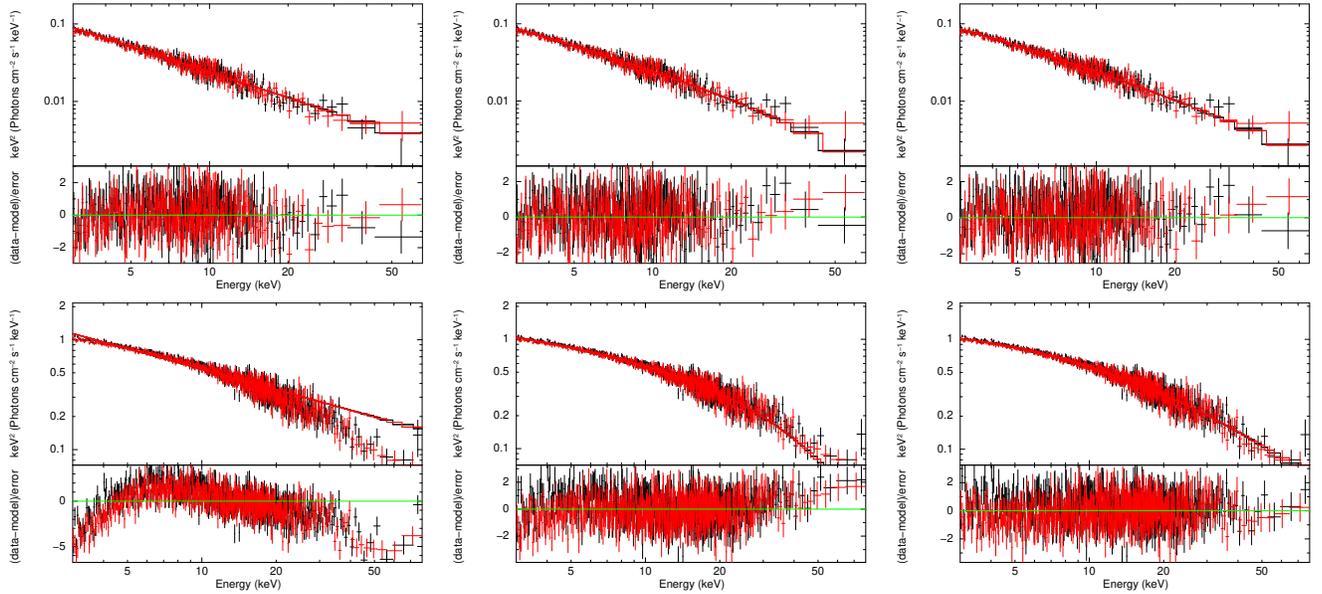

\begin{center}
\includegraphics[width=0.22\linewidth, angle =270]{12-nu-pl.eps}
\includegraphics[width=0.22\linewidth, angle =270]{12-nu-cpl.eps}
\includegraphics[width=0.22\linewidth, angle =270]{12-nu-lp.eps}
\includegraphics[width=0.22\linewidth, angle =270]{19-nu-pl.eps}
\includegraphics[width=0.22\linewidth, angle =270]{19-nu-cpl.eps}
\includegraphics[width=0.22\linewidth, angle =270]{19-nu-lp.eps}
\caption {Spectral fits (\emph{NuSTAR} alone) using the models PL, CPL and LP 
(left to right) for the ObsIDs 60002023018	 (low-flux state) and 60002023027 (high-flux 
state) are shown in the upper and lower panels, respectively. Significant curvature is apparent in the high flux state.}
\label{fig:nu-pl-lp-cutoff-fit}
\end{center}
\end{figure*}
\begin{figure*}
\includegraphics[width=0.45\textwidth]{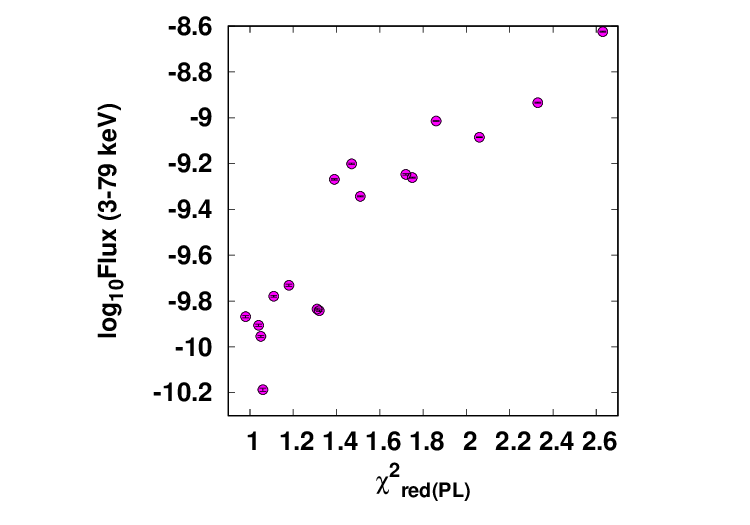}
\includegraphics[width=0.45\textwidth]{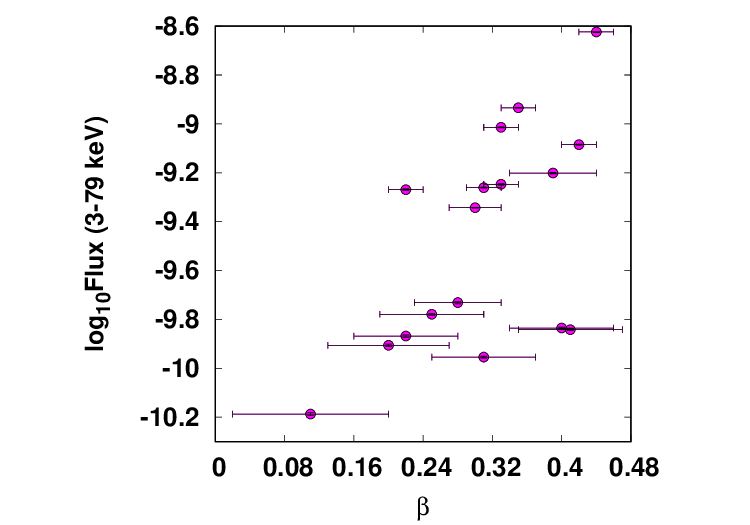}
\caption {Scatter plots (\emph{NuSTAR} alone) showing flux, along with reduced chi-square values for the PL fit (left), and LP curvature values $\beta$ (right).}
\label{fig:flux-var-f}
\end{figure*}
\begin{figure*}
\includegraphics[width=0.32\textwidth]{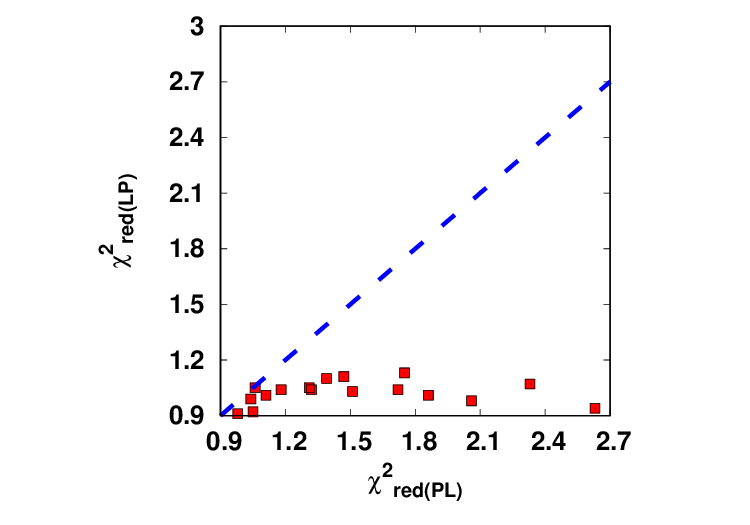}
\includegraphics[width=0.32\textwidth]{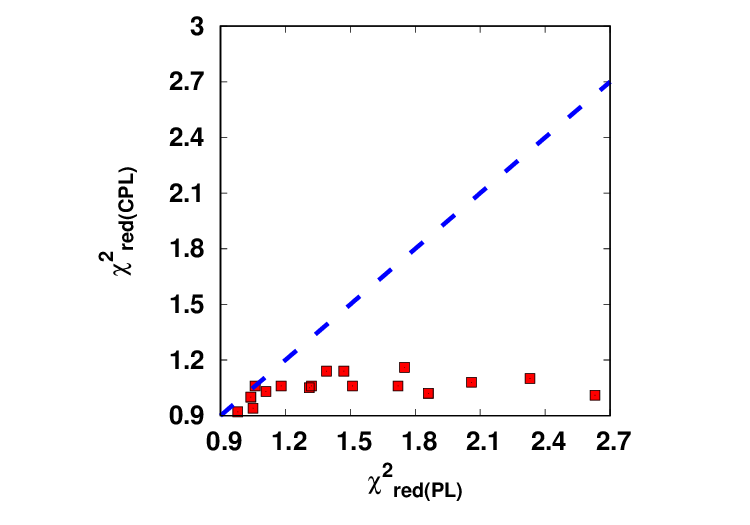}
\includegraphics[width=0.32\textwidth]{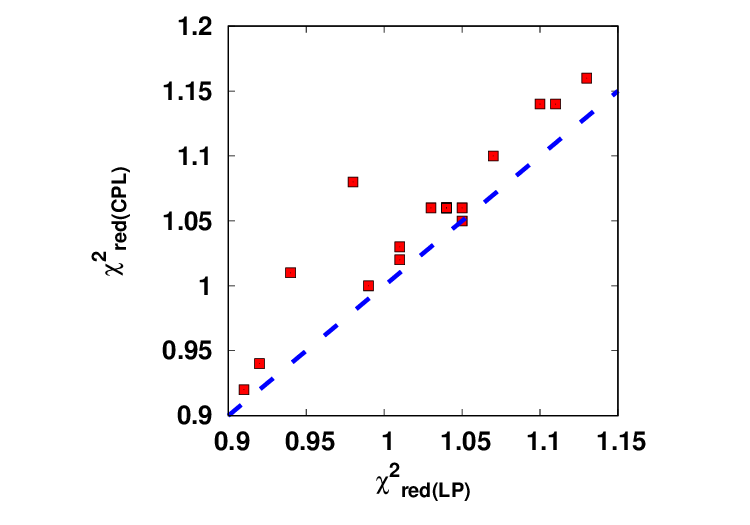}
\caption{Scatter plots between the reduced chi-square values of \emph{NuSTAR} data fitted with the PL, LP, and CPL models, along with the identity line. }
\label{fig:chisq-plots}
\end{figure*}
\begin{figure*}
\includegraphics[width=0.45\textwidth]{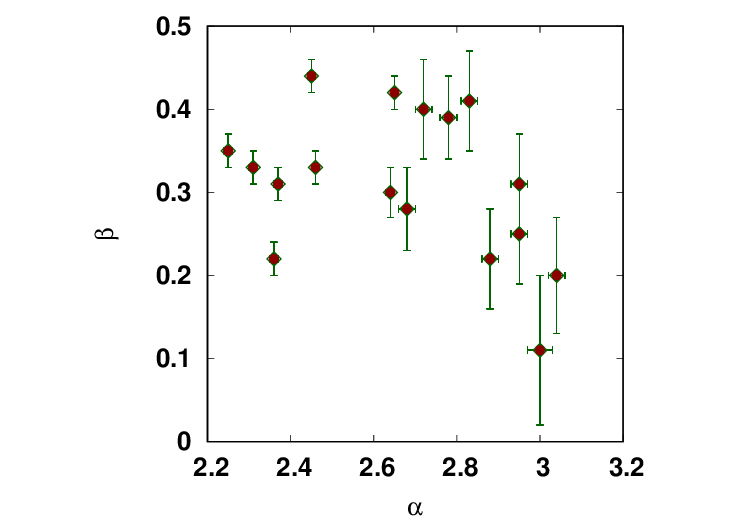}
\includegraphics[width=0.45\textwidth]{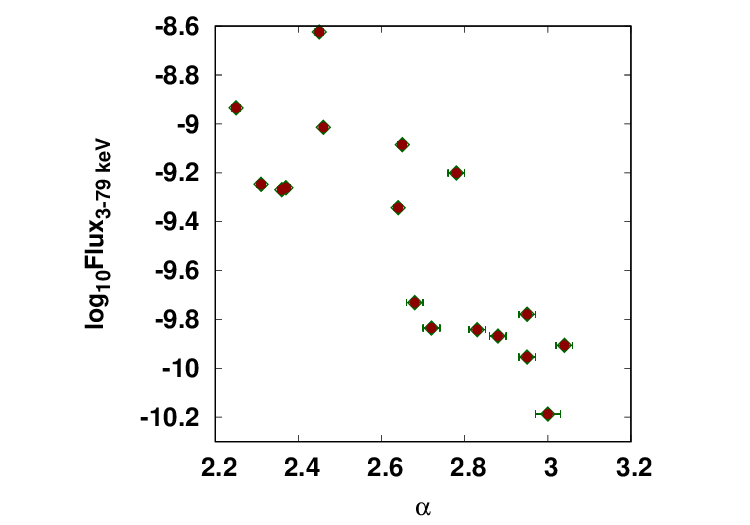}
\caption {Scatter plots (\emph{NuSTAR} alone) showing LP index $\alpha$, along with 
spectral curvature $\beta$ (left), and flux in the 3-79 keV range (right).}
\label{fig:nu-logpar}
\end{figure*}
\begin{figure*}
\includegraphics[width=0.32\textwidth]{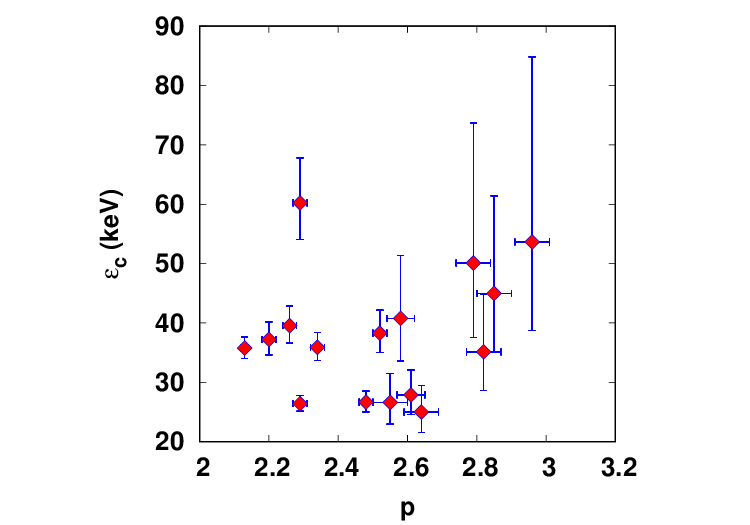}
\includegraphics[width=0.32\textwidth]{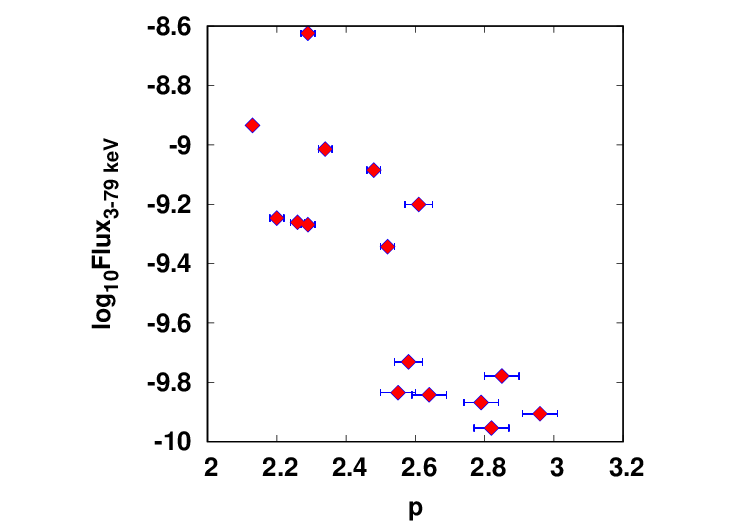}
\includegraphics[width=0.32\textwidth]{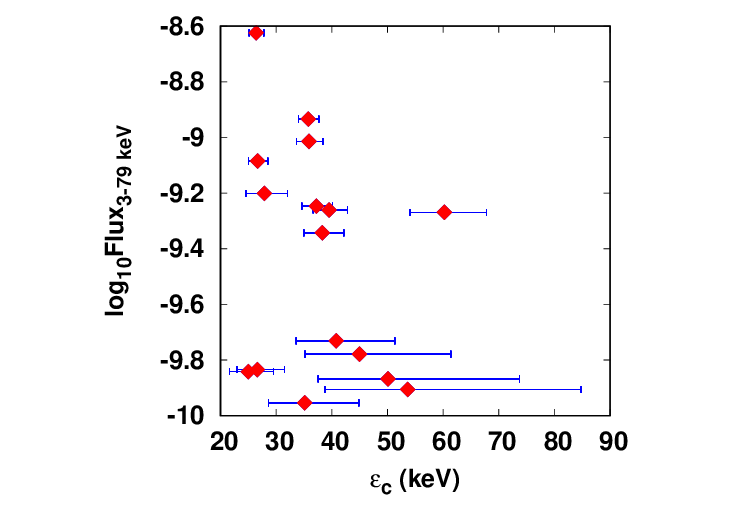}
\caption{Scatter plots (\emph{NuSTAR} alone) showing best-fit CPL model parameters: index and cutoff energy (in keV), index and flux, and cutoff energy and flux (from left to right).} 
\label{fig:nu-cutoffpl}
\end{figure*}
\begin{table*}
\begin{center}
\caption{ The best-fit parameters of combined \emph{Swift}-XRT and \emph{NuSTAR} spectral fitting 
with a LP and a simple CPL model, respectively.} 
\begin{tabular}{cc c c c c c c c c c c } 
\hline
\multicolumn{2}{c}{Obs.ID} 	&	Flux	&&	\multicolumn{4}{c} {LP ($\epsilon_0$=5 keV)}				&&			\multicolumn{3}{c} {CPL}				\\
\cline{5-8}
\cline{10-12}
\emph{Swift}	&	\emph{\emph{NuSTAR}}&		(0.3-79 keV)	&&	$\alpha$			&	$\beta$			& $\epsilon_p$ (keV)&	$\chi_{red}^2$(dof)	&&	$p$				&	$\epsilon_c$ (keV)			&	$\chi_{red}^2$(dof)				\\

\hline
35014034	&	60002023006	&	-9.696$\pm$0.003	&	&	2.93$\pm$0.01	&	0.34$\pm$0.01	&	0.22$\pm$0.02	&	0.97	(962)&	&	2.38$\pm$0.01	&	11.5$\pm$0.42	&	1.17	(962)	\\
80050003	&	60002023010	&	-9.602$\pm$0.003	&	&	2.81$\pm$0.01	&	0.36$\pm$0.01	&	0.37$\pm$0.02	&	1.27	(1110)&	&	2.23$\pm$0.01	&	10.37$\pm$0.3	&	1.4	(1110)	\\
80050006	&	60002023014	&	-9.93$\pm$0.004	&	&	2.94$\pm$0.02	&	0.25$\pm$0.01	&	0.07$\pm$0.01	&	1.19	(848)&	&	2.51$\pm$0.01	&	14.4$\pm$0.87	&	1.3	(848)	\\
80050007	&	60002023016	&	-9.523$\pm$0.003	&	&	2.92$\pm$0.01	&	0.32$\pm$0.02	&	0.19$\pm$0.03	&	1.1	(822)&	&	2.46$\pm$0.02	&	13.71$\pm$0.72	&	1.34	(822)	\\
80050011	&	60002023018	&	-9.639$\pm$0.003	&	&	2.98$\pm$0.01	&	0.34$\pm$0.01	&	0.17$\pm$0.01	&	1.1	(1001)&	&	2.42$\pm$0.01	&	10.73$\pm$0.38	&	1.35	(1001)	\\
80050013	&	60002023020	&	-9.521$\pm$0.003	&	&	2.67$\pm$0.01	&	0.24$\pm$0.01	&	0.19$\pm$0.02	&	1.16	(1126)&	&	2.29$\pm$0.01	&	16.97$\pm$0.69	&	1.2	(1126)	\\
80050014	&	60002023022	&	-9.146$\pm$0.002	&	&	2.64$\pm$0.01	&	0.31$\pm$0.01	&	0.47$\pm$0.03	&	1.06	(1338)&	&	2.23$\pm$0.01	&	16.45$\pm$0.46	&	1.62	(1338)	\\
80050016	&	60002023024	&	-8.972$\pm$0.002	&	&	2.78$\pm$0.01	&	0.36$\pm$0.01	&	0.41$\pm$0.02	&	1.18	(1039)&	&	2.23$\pm$0.01	&	11.93$\pm$0.38	&	1.41	(1039)	\\
80050019	&	60002023027	&	-8.46$\pm$0.001	&	&	2.45$\pm$0.01	&	0.43$\pm$0.01	&	1.5$\pm$0.03	&	1.07	(1575)&	&	1.91$\pm$1.91	&	12.16$\pm$12.16	&	2.11	(1575)	\\
32792002	&	60002023029	&	-8.874$\pm$0.001	&	&	2.68$\pm$0.01	&	0.34$\pm$0.01	&	0.51$\pm$0.02	&	1.1	(1411)&	&	2.21$\pm$0.01	&	14.15$\pm$0.31	&	1.52	(1411)	\\
35014062	&	60002023033	&	-8.847$\pm$0.002	&	&	2.47$\pm$0.01	&	0.29$\pm$0.01	&	0.79$\pm$0.05	&	1.07	(1391)&	&	2.19$\pm$0.01	&	22.72$\pm$0.83	&	1.33	(1391)	\\
35014065	&	60002023035	&	-8.799$\pm$0.001	&	&	2.27$\pm$0.01	&	0.28$\pm$0.01	&	1.61$\pm$0.05	&	1.17	(1741)&	&	1.96$\pm$0.01	&	21.77$\pm$0.52	&	1.45	(1741)	\\
35014066	&	60002023037	&	-9.6$\pm$0.003	&	&	2.76$\pm$0.01	&	0.22$\pm$0.01	&	0.1$\pm$0.01	&	1.19	(1068)&	&	2.39$\pm$0.01	&	17.67$\pm$0.77	&	1.14	(1068)	\\
35014067	&	60002023039	&	-9.625$\pm$0.003	&	&	2.85$\pm$0.01	&	0.26$\pm$0.01	&	0.11$\pm$0.01	&	1.05	(1001)&	&	2.43$\pm$0.01	&	14.71$\pm$0.65	&	1.21	(1001)	\\
34228110	&	60202048002	&	-9.133$\pm$0.002	&	&	2.34$\pm$0.01	&	0.26$\pm$0.01	&	1.14$\pm$0.08	&	1.09	(1348)&	&	2.12$\pm$0.01	&	28.88$\pm$1.44	&	1.5	(1348)	\\
81926001	&	60202048004	&	-9.105$\pm$0.002	&	&	2.33$\pm$0.01	&	0.29$\pm$0.01	&	1.33$\pm$0.08	&	1.07	(1385)&	&	2.05$\pm$0.01	&	23.45$\pm$0.92	&	1.29	(1385)	\\
34228145	&	60202048006	&	-9.116$\pm$0.003	&	&	2.37$\pm$0.01	&	0.31$\pm$0.02	&	1.29$\pm$0.15	&	1.11	(1044)&	&	2.25$\pm$0.02	&	37.69$\pm$2.64	&	1.18	(1044)	\\
	
\hline
\end{tabular}  
\label{tab:NuSTAR-simul-fit-det}
\end{center}   
\end{table*}

\begin{figure*}
\begin{center}
\includegraphics[width=0.32\linewidth]{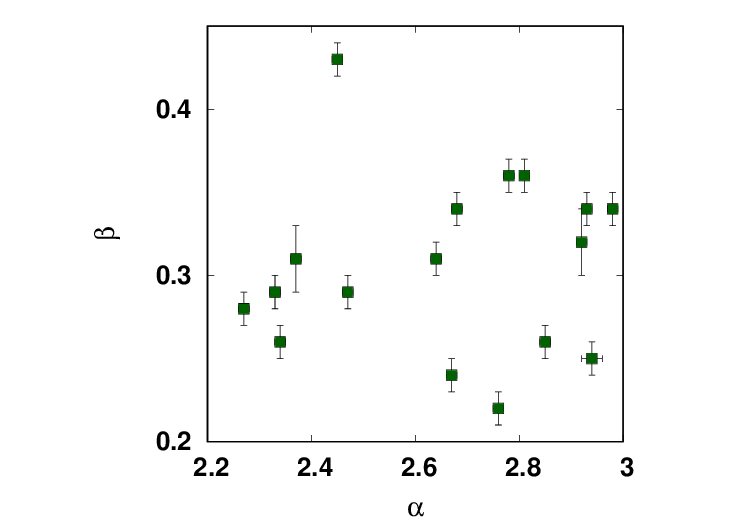}
\includegraphics[width=0.32\linewidth]{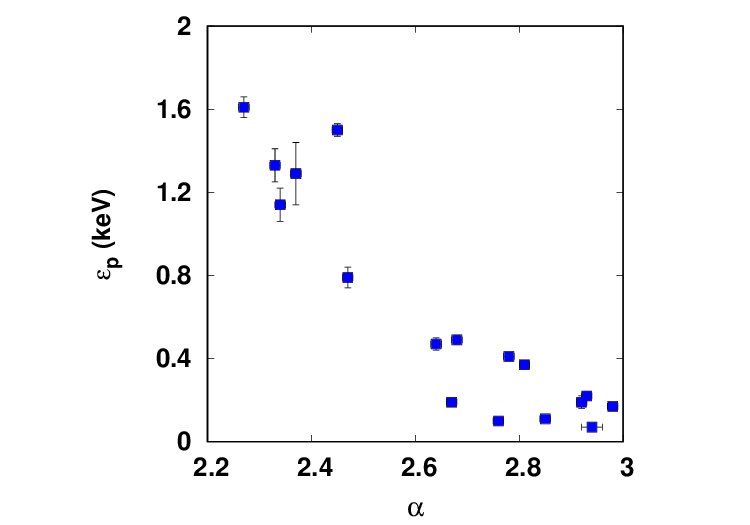}
\includegraphics[width=0.32\linewidth]{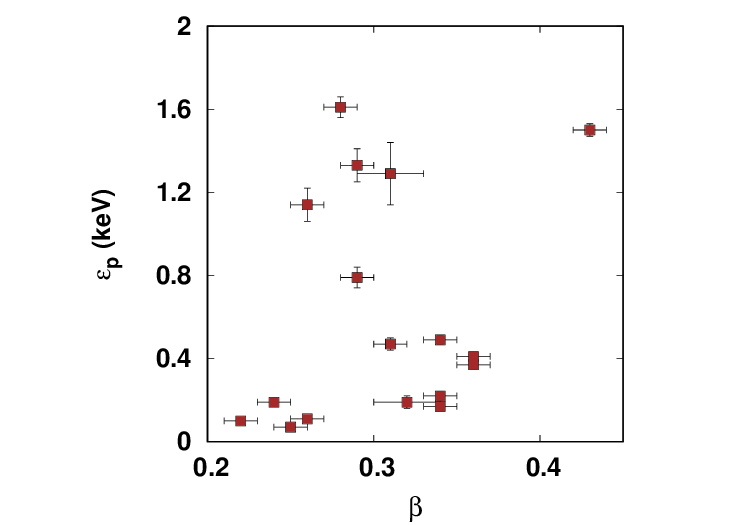}
\includegraphics[width=0.32\linewidth]{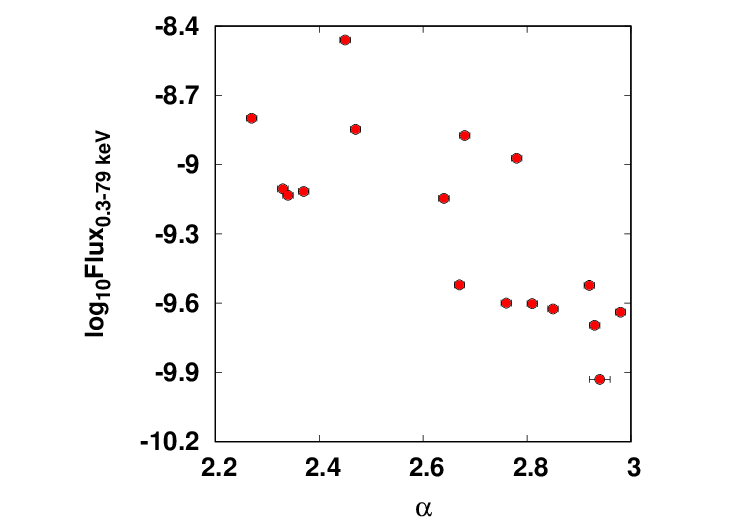}
\includegraphics[width=0.32\linewidth]{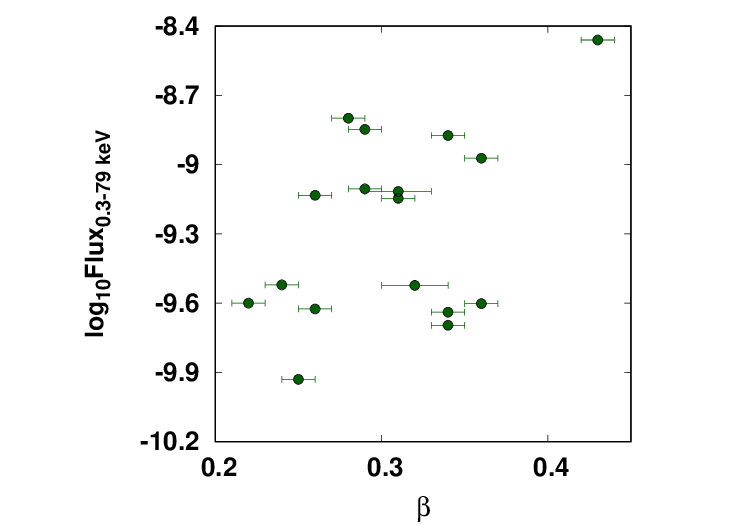}
\includegraphics[width=0.32\linewidth]{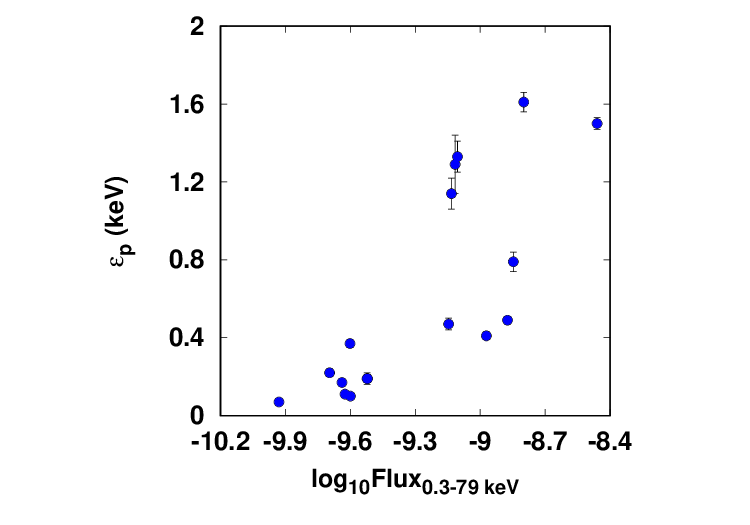}
\caption {Scatter plots between the parameters obtained from LP fitting of 
combined \emph{Swift}-XRT and \emph{NuSTAR} data. The upper panel represents plots between index ($\alpha$) and curvature ($\beta$), $\alpha$ and peak energy ($\epsilon_p$), and $\beta$ and $\epsilon_p$ (from left to right). The lower panel shows the variation of flux in 0.3 to 79 keV with $\alpha$, $\beta$ and $\epsilon_p$, respectively (from left to right).}
\label{fig:combfit-fit-plots}
\end{center}
\end{figure*}

The neutral hydrogen column density, inclusive of both HI and HII was fixed at $N_H = 2.03\times 10^{20} \rm cm^{-2}$ throughout the analysis. The best-fit parameters of these models 
are presented in Table \ref{tab:nustar-alone-fit}. We observe that most of the low-flux 
states are well-fitted with a steep power-law model, yielding a photon index 
saturating at $\sim$ 3, as reported earlier \cite{Balokovi__2016}. In contrast, the 
high-flux states exhibit significant curvature and deviate from a simple power-law model 
(see Figure \ref{fig:nu-pl-lp-cutoff-fit}). The spectral fittings using the PL, LP, and CPL models 
for a sample of low and high flux states are shown in Figure \ref{fig:nu-pl-lp-cutoff-fit}. 
Figure \ref{fig:flux-var-f} illustrates the change in the reduced chi-square value of 
the PL fit (left) and the spectral curvature (right) as the flux increases.\\
Our analysis provides strong evidence for spectral curvature in the \emph{NuSTAR} regime, with a 
LP/CPL model clearly preferred over a pure PL model in high flux states. The plots between 
reduced chi-square values for the spectral fittings with the PL, LP, and CPL models 
($\chi_{\rm red}^2(\rm PL)$, $\chi_{\rm red}^2(\rm LP)$, and $\chi_{\rm red}^2(\rm CPL)$) 
are shown in Figure \ref{fig:chisq-plots}. Scatter plots showing the best-fit LP and CPL 
model parameters, and with the flux, are depicted in Figures \ref{fig:nu-logpar} and 
\ref{fig:nu-cutoffpl}, respectively. To identify the dependence between these best-fit 
parameters, we performed a Spearman rank correlation analysis between various quantities 
obtained from the spectral fit. The Spearman rank correlation study between the LP model 
parameters $\alpha$ and $\beta$ yields a correlation coefficient, $r_s$= -0.41 with a 
null hypothesis probability, $p=0.104$. This result is consistent with previous studies 
which reported that no significant correlation was observed \citep{2015A&A...580A.100S,10.1093/mnras/sty2003}. 
However, an anti-correlation is witnessed between $\alpha$ and the flux ($r_s$ = -0.81, 
p $<$0.001), indicating that the spectra get harder during brighter states of the source. 
This harder when brighter behaviour of the source has already been reported earlier 
\citep{2004A&A...413..489M, 2008A&A...478..395M,  2015A&A...580A.100S, 2018ApJ...858...68K, 
Kapanadze_2020}. The spectral fittings with the simple CPL model allows to  constrain well
the cutoff energy during high-flux states. Here no correlation is observed between 
$p$ and $\epsilon_c$ ($r_s$ = 0.16, p = 0.556), while a significant anti-correlation between 
$p$ and flux ($r_s$ = -0.79 p $<$0.001) is seen.\\

\subsection{Combined \emph{NuSTAR} and \emph{Swift}-XRT (0.3-79 keV) regime}
In order to gain further insights, we have also studied the broad X-ray spectra of Mkn 421 
ranging from 0.3 to 79 keV using simultaneous \emph{Swift}-XRT and \emph{NuSTAR} observations, employing 
the log-parabola (LP) and simple exponential cutoff (CPL) model, respectively. These 
X-ray spectra exhibits significant curvature, and a LP model generally provides a better 
fit when compared to a simple CPL model. The best-fit parameters are presented in Table 
\ref{tab:NuSTAR-simul-fit-det}, and the scatter plots between LP parameters and the flux 
are shown in Figure \ref{fig:combfit-fit-plots}. Again, we observe no correlation between 
the $\alpha$ and $\beta$ ($r_s$  = 0.17, p = 0.521). 
Additionally, there was no significant correlation between $\beta$ and peak energy, $\epsilon_p$,  
($r_s$ = 0.31, p = 0.227), whereas $\alpha$ showed a strong negative correlation with 
$\epsilon_p$ ($r_s$ = -0.86, p $<$0.001). Furthermore, we noted a strong negative correlation 
between $\alpha$ and flux ($r_s$ = -0.80, p $<$0.001), and $\epsilon_p$ being significantly 
correlated with flux ($r_s$ = 0.85, p $<$0.001). These correlations suggest that during 
flares, the spectral index hardens and the spectral peak moves towards higher 
energies.

The absence of a significant correlation among the LP parameters, even in the broad energy 
range studied here, indicates that the changes in spectral characteristics cannot simply 
be ascribed to the energy-dependence of the particle acceleration process as proposed in 
ref.~\citep{2004A&A...413..489M}. Additionally, such a model is unable to account for the
broadband SED of blazars \citep{2004A&A...413..489M,2009A&A...501..879T,2015A&A...580A.100S}. 
Hence, an alternate physically motivated choice could be a CPL type model. On the other 
hand, the foregoing analysis indicates that a simple (purely exponential) CPL model does 
not provide a better fit to the broad X-ray spectra of the source, particularly in high 
flux states. To explore this further, we next study the broadband X-ray spectrum using a power-law with a modified exponential cutoff, as might be expected to occur in shock-type acceleration scenarios (see Section \S \ref{sec_intro}).

\begin{table*}
\begin{center}
\caption{ Best fit parameters using the MCPL model (cooled $p$=2) for the energy range $\epsilon_p$-79 keV.}
\begin{tabular}{c c c c c c c } 
\hline
\multicolumn{2}{c}{Obs.ID} &	$\epsilon_p$			&	$\epsilon_c$		&			$\zeta$				&	$\chi_{red}^2$(dof)				&		Flux\\
\emph{Swift}	&	\emph{NuSTAR} &	(keV)				&	(keV)	&		&		&						($\epsilon_p$-79 keV)				\\
\hline
80050003	&	60002023010	&	0.37$\pm$0.02	&	2.64$\pm$0.16	&	0.54$\pm$0.01	&	1.2	(1106)	&	-9.621$\pm$0.003	\\
80050014	&	60002023022	&	0.47$\pm$0.03	&	2.95$\pm$0.32	&	0.47$\pm$0.02	&	1.17	(1323)	&	-9.156$\pm$0.002	\\
80050016	&	60002023024	&	0.41$\pm$0.02	&	2.63$\pm$0.27	&	0.52$\pm$0.02	&	1.18	(1030)	&	-8.988$\pm$0.003	\\
80050019	&	60002023027	&	1.5$\pm$0.03	&	8.01$\pm$0.53	&	0.6$\pm$0.02	&	1.03	(1458)	&	-8.457$\pm$0.002	\\
32792002	&	60002023029	&	0.49$\pm$0.02	&	3.54$\pm$0.28	&	0.52$\pm$0.02	&	1.1	(1394)	&	-8.888$\pm$0.002	\\
35014062	&	60002023033	&	0.79$\pm$0.05	&	6.9$\pm$0.66	&	0.53$\pm$0.02	&	1.09	(1345)	&	-8.853$\pm$0.002	\\
35014065	&	60002023035	&	1.61$\pm$0.05	&	20.58$\pm$0.66	&	0.71$\pm$0.03	&	1.05	(1613)	&	-8.807$\pm$0.002	\\
34228110	&	60202048002	&	1.14$\pm$0.08	&	7.9$\pm$1.54	&	0.44$\pm$0.04	&	1.11	(1266)	&	-9.128$\pm$0.003	\\
81926001	&	60202048004	&	1.33$\pm$0.08	&	14.96$\pm$1.01	&	0.62$\pm$0.04	&	1.06	(1285)	&	-9.109$\pm$0.003	\\
34228145	&	60202048006	&	1.29$\pm$0.15	&	8.88$\pm$1.18	&	0.5$\pm$0.03	&	1.12	(1015)	&	-9.112$\pm$0.003	\\

\hline
\end{tabular} 
\label{tab:semiexpcut-fit}  
 \end{center}   
 \end{table*}
\begin{table*}
\begin{center}
\caption{ Best fit parameters using the MCPL model by assuming Bohm($\zeta$=0.5) and Hard-sphere($\zeta$=0.33) for 
$\epsilon_p$-79 keV fit.} 
 \begin{tabular}{c cc c c c c c c c c } 
\hline
\multicolumn{2}{c}{Obs.ID} &	$\epsilon_p$			&	\multicolumn{3}{c}{Bohm: $\zeta$=0.5}&&\multicolumn{3}{c}{Hard-sphere: $\zeta$=0.33}\\
\cline{4-6}
\cline{8-10}
\emph{Swift}	&	\emph{NuSTAR}		&		(keV)		&	$p$		&			$\epsilon_c$			&	$\chi_{red}^2$(dof)				&&		$p$			&		$\epsilon_c$				&		$\chi_{red}^2$(dof)				\\
\hline
80050003	&	60002023010	&	0.37$\pm$0.02	&	1.95$\pm$0.02	&	1.84$\pm$0.1	&	1.2	(1106)&&	1.63$\pm$0.02	&	0.11$\pm$0.01	&	1.19	(1106)\\
80050014	&	60002023022	&	0.47$\pm$0.03	&	2$\pm$0.02	&	3.6$\pm$0.22	&	1.18	(1323)&&	1.72$\pm$0.03	&	0.25$\pm$0.02	&	1.12	(1323)\\
80050016	&	60002023024	&	0.41$\pm$0.02	&	1.97$\pm$0.02	&	2.2$\pm$0.14	&	1.17	(1030)&&	1.65$\pm$0.03	&	0.13$\pm$0.01	&	1.16	(1030)\\
80050019	&	60002023027	&	1.5$\pm$0.03	&	1.83$\pm$0.03	&	3.33$\pm$0.24	&	1.01	(1457)&&	1.43$\pm$0.04	&	0.16$\pm$0.02	&	0.98	(1457)\\
32792002	&	60002023029	&	0.49$\pm$0.02	&	1.95$\pm$0.02	&	2.75$\pm$0.13	&	1.1	(1394)&&	1.64$\pm$0.03	&	0.17$\pm$0.01	&	1.06	(1394)\\
35014062	&	60002023033	&	0.79$\pm$0.05	&	1.93$\pm$0.03	&	4.83$\pm$0.41	&	1.08	(1345)&&	1.6$\pm$0.04	&	0.3$\pm$0.04	&	1.05	(1345)\\
35014065	&	60002023035	&	1.61$\pm$0.05	&	1.8$\pm$0.02	&	6.27$\pm$0.51	&	1.03	(1613)&&	1.49$\pm$0.04	&	0.41$\pm$0.05	&	1.03	(1613)\\
34228110	&	60202048002	&	1.14$\pm$0.08	&	2.04$\pm$0.03	&	13.05$\pm$1.97	&	1.11	(1266)&&	1.82$\pm$0.05	&	1.19$\pm$0.25	&	1.1	(1266)\\
81926001	&	60202048004	&	1.33$\pm$0.08	&	1.86$\pm$0.03	&	6.24$\pm$0.7	&	1.05	(1284)&&	1.55$\pm$0.05	&	0.4$\pm$0.06	&	1.04	(1284)\\
34228145	&	60202048006	&	1.29$\pm$0.15	&	1.99$\pm$0.04	&	8.32$\pm$1.1	&	1.12	(1015)&&	1.71$\pm$0.06	&	0.6$\pm$0.11	&	1.12	(1015)\\

\hline
\end{tabular}
\label{table:hard-bohm}  
\end{center}   
\end{table*}

 \begin{figure*}
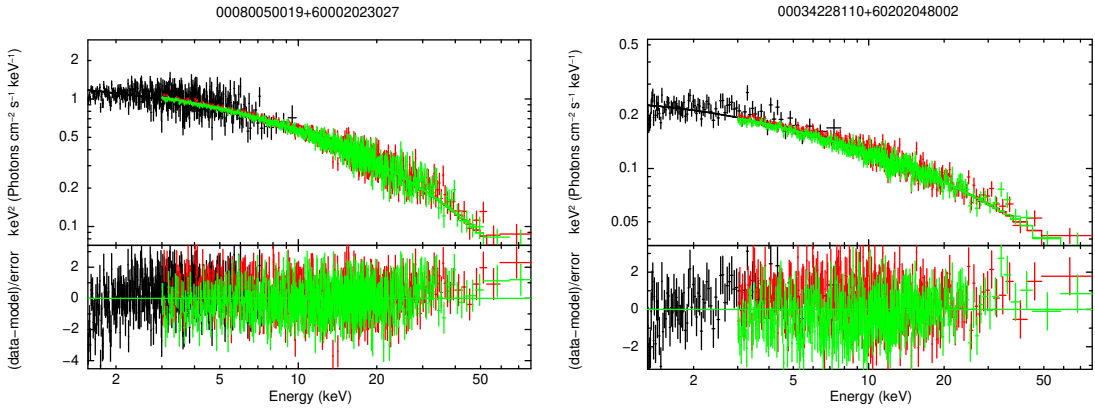

\begin{center}
\includegraphics[width=0.3\linewidth, angle =270]{19-semiexpcut.eps}
\includegraphics[width=0.3\linewidth, angle =270]{26-semiexpcut.eps}
\caption {Spectral fit from  $\epsilon_p$-79 keV using the MCPL model for the obsID 
00080050019+60002023027 (left) and  00034228110+60202048002 (right).}
\label{fig:comb-semiexpcut-fit-spec}  
\end{center}
\end{figure*}

 \begin{figure*}
\begin{center}
\includegraphics[width=0.45\linewidth]{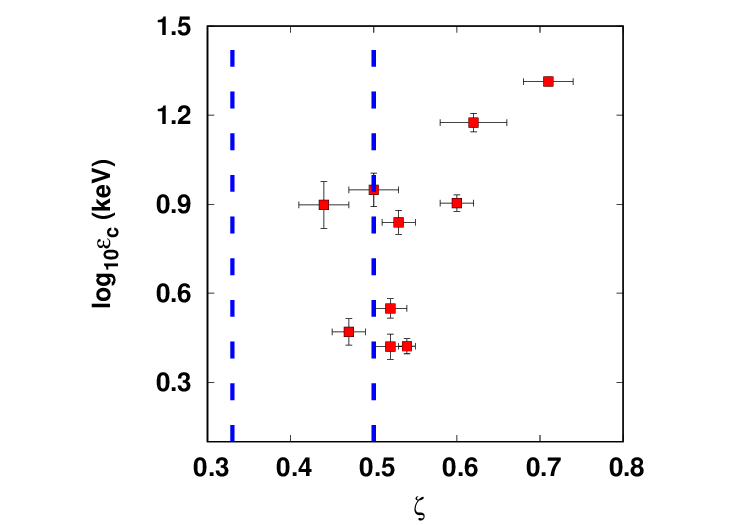}
\includegraphics[width=0.45\linewidth]{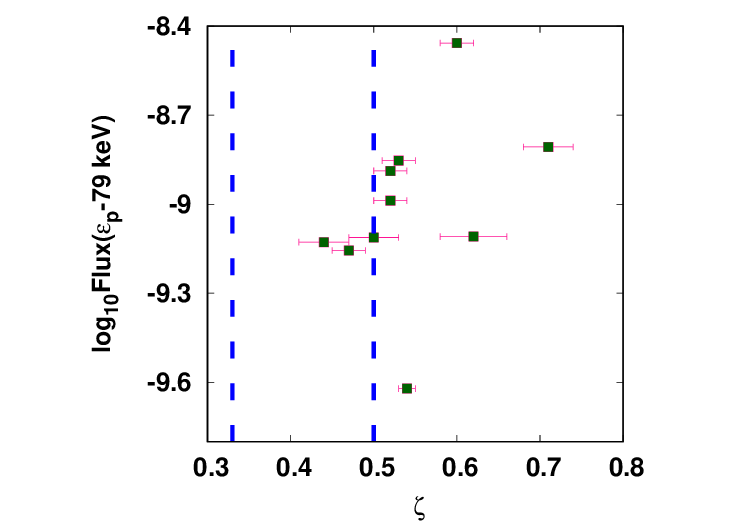}
\includegraphics[width=0.45\linewidth]{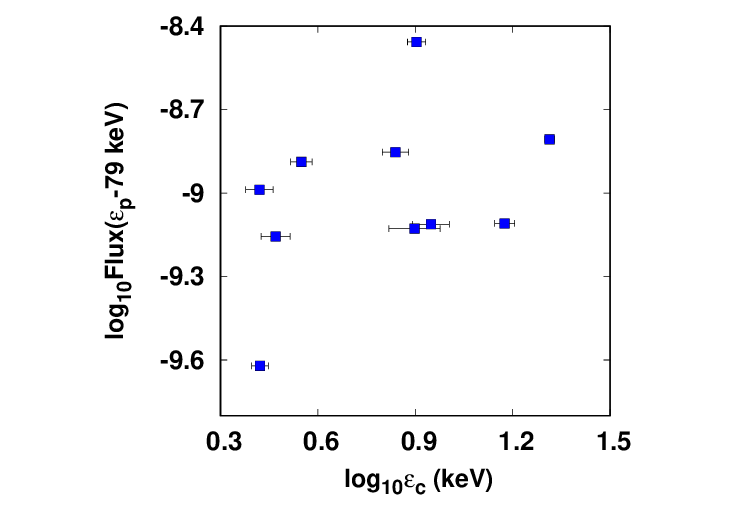}
\caption {Scatter plots between best-fit parameters for the MCPL model 
($\epsilon_p$-79 keV). The upper left panel is for $\zeta$ and cutoff energy ($\epsilon_c$), 
 right is for $\zeta$ and flux in $\epsilon_p$-79 keV, and the lower panel is for $\epsilon_c$ and flux in $\epsilon_p$-79 keV. The dotted vertical lines represent $\zeta$ corresponding to hard-sphere (0.33) and Bohm (0.5).}
\label{fig:comb-semiexpcut-plots-index-2} 
\end{center}
\end{figure*}

\subsection{Probing a power-law with modified exponential cutoff}
In the context of shock acceleration scenarios, the electron distribution exhibits a simple 
exponential cutoff form only when diffusion is independent of energy. The shape of the 
particle distribution deviates from this as the diffusion coefficient becomes energy-dependent, 
leading to a corresponding change in the synchrotron cutoff (i.e., typically sub-exponential) 
shape \citep{Zirakashvili_2007,1989A&A...214...14F}.
In HBL sources such as Mkn 421, the maximum achievable electron energies are limited by 
synchrotron losses.
The spectral index evolution around the synchrotron (SED) peak does not support a 
simple cooling break origin of the peak frequency, rather might be affected by the
blending of different components \citep[e.g.,][]{2022MNRAS.514.3074B}. 
At hard X-rays, where the synchrotron spectrum declines, we may expect the emission to be 
more dominated by a single component, particularly during higher flux states. Additionally, 
the high-energy end of the spectrum is likely to be populated by a cooled electron distribution. 
To explore the spectral curvature towards high energies in more detail, we thus perform 
spectral fits to the X-ray data above the synchrotron SED peak (with $\epsilon_p$ $>$ 0.3 keV) 
using a modified CPL (MCPL) model

\begin{equation}
   F(\epsilon) \propto \epsilon^{-p}\, \rm exp[-(\epsilon/\epsilon_c)^\zeta] 
    \quad\quad \mathrm{(MCPL)}\,,
\end{equation}
where $p$ represents the power-law index, $\epsilon_c$ characterizes the position of the cutoff 
energy, and the parameter $\zeta$ governs the steepness of the cutoff. This function is added 
as a local model in XSPEC, and we perform spectral fitting for the combined simultaneous 
\emph{Swift}-XRT and \emph{NuSTAR} observations from $\epsilon_p$ to 79 keV. 

 Among the total 17 simultaneous \emph{Swift}-XRT and \emph{NuSTAR} observations we found only for 10 epochs, the peak falls in between 0.3--79 keV. Most of these epochs are during high-flux states (except obsID. 80050003+60002023010), and all these epochs are considered for the MCPL fit. The considered 
X-ray spectra did not allow us to constrain all parameters of the model. Hence, 
we performed a fitting with $p$ fixed to a value 2, representing a cooled particle distribution. 
This choice may be appropriate since we are interested in the spectrum above $\epsilon_p$ where 
synchrotron losses dominate.  The sample spectral fits of MCPL model are shown in Figure \ref{fig:comb-semiexpcut-fit-spec}.
The modified CPL model  represents well the spectrum above $\epsilon_p$ and the best-fit parameters 
$\epsilon_c$ and $\zeta$ are shown in Table \ref{tab:semiexpcut-fit}. The scatter plots between 
the fitting parameters, and with the flux are shown in Figure \ref{fig:comb-semiexpcut-plots-index-2}.  
 We performed a Spearman correlation analysis and did not find a significant correlation between the MCPL model
parameters $\epsilon_c$ and $\zeta$ ($r_s$ = 0.45, $p$ = 0.192). Also, no significant correlations are observed between $\epsilon_c$  and flux ($r_s$ = 0.42, $p$ = 0.229), and  $\zeta$ and flux ($r_s$ = 0.55, $p$ = 0.102).

Constraining the $\zeta$-parameter in the X-ray spectrum can provide insights into the parent 
particle distribution. In the case of synchrotron emission, the parameter $\zeta$ is linked to 
the primary particle distribution through the relation $\zeta = \frac{\beta_e}{\beta_e+2}$ 
\citep{1989A&A...214...14F}. Therefore, the value of $\zeta$ is expected to be 0.33 in the 
case of energy-independent diffusion ($\beta_e=1$), while Bohm-type diffusion ($\beta_e =2$)
results in $\zeta$=0.5. As can be seen from Table~\ref{tab:semiexpcut-fit}, the inferred 
$\zeta$-values favour a Bohm-type behaviour. 

For comparison, we also repeated this analysis by fixing $\zeta$ at the values corresponding 
to hard-sphere and Bohm-type diffusion, which are 0.33 and 0.5, respectively (Table \ref{table:hard-bohm}). The best-fit 
parameters revealed that $\zeta =0.5$ leads to an index $p$ closer to $\sim$ 2, supporting our 
previous assumption of a cooled distribution above $\epsilon_p$. 
Therefore, the results appear consistent with a cooled particle distribution with a cutoff 
shaped by Bohm-type diffusion ($\beta_e$ = 2).

\section{Summary}
\label{sec_discussion}
We have conducted a detailed study of the X-ray spectra of Mkn 421 using simultaneous 
\emph{Swift}-XRT and \emph{NuSTAR} observations. Most of our observations considered are during 
high flux states (unlike ref.~\cite{Balokovi__2016} for example) and 
enables us to investigate the spectral curvature in the hard X-ray regime more rigorously. 
Our spectral study of \emph{NuSTAR} observations using power-law (PL), log-parabolic (LP), and
simple exponential cutoff power-law (CPL) models suggests that LP and CPL are clearly 
preferred over a simple PL. This provides strong evidence of spectral curvature in the 
3-79 keV energy regime. We also examined the broad (\emph{Swift}-XRT and \emph{NuSTAR}) X-ray spectra, 
spanning from 0.3 to 79 keV with LP and CPL models, indicating that LP provides a better 
fit compared to CPL. However, the lack of a significant correlation between the LP 
parameters suggests that the variations in spectral characteristics cannot be 
attributed to the energy-dependence of the particle acceleration process. 

The curvature in the X-ray spectrum is closely linked to the primary electron distribution. 
The acceleration of electrons at shocks is a favored mechanism for generating non-thermal 
particle distributions in astrophysical jets. In the presence of radiative losses like the
synchrotron process, the accelerated electron distribution will be a broken power-law with 
a modified exponential cutoff at the maximum available electron energy. The resultant 
synchrotron spectrum from such a particle distribution will always be a power-law with 
a sub-exponential cutoff, and we found that a MCPL function can satisfactorily reproduce 
the data beyond the SED peak. Further, the results are consistent with a scenario where 
the hard X-ray spectrum is due to a cooled electron distribution, with the highest energy 
part shaped by Bohm-type diffusion.
For a strong shock that is non-relativistic in the jet frame ($\Gamma_s = \Gamma_j
\Gamma_b (1-\beta_b\beta_j)\sim 1$, with $\Gamma_j\gg1$ being the jet Lorentz factor 
and $\Gamma_b$ the `blob' Lorentz factor), the characteristic acceleration timescale 
is approximately given by $t_{\rm acc}' \simeq 10 \kappa'/u_s^2$ where $\kappa' =(1/3)\, 
\lambda' c$ and $\lambda' \sim r_g' = \gamma_e' m_c c^2/(eB')$ in the Bohm limit 
\citep[e.g.,][]{Rieger_2007}. Balancing acceleration with cooling, $t_{\rm syn}'=
9 m_e^3c^5/(4e^4\gamma_e' B'^2)$, one can estimate maximum achievable electron 
energies ($\gamma_{\rm e,m}'$). The corresponding synchrotron photon energy $\epsilon_c' 
\propto \gamma_{e,m}'^2 B'$ can be compared to the synchrotron cutoff energies $\epsilon_c 
\sim 10$ keV inferred from observations (Table~\ref{tab:semiexpcut-fit}) taking beaming 
($\epsilon_c \sim \Gamma_j \epsilon_c'$) into account. The result then substantiates the 
initial assumption of non-relativistic shock acceleration.
 
Though the observed spectral curvature in the hard X-ray supports Bohm-type diffusion 
during flares in the jet, other possibilities are not yet be ruled out. For instance, 
an energy-dependent escape time scale or the superposition of multiple broken power-law 
components might also contribute to spectral curvature. Since the highest-energy 
electrons are expected to probe the shock vicinity, X-ray spectral analysis along with 
dedicated polarization studies \cite[e.g.,][]{DiGesu2022} provides a powerful diagnostics 
of the underlying flow properties. The signature of an electron distribution shaped by 
Bohm-type diffusion process could in principle be further probed by modeling the resultant 
$\gamma$-ray emission by inverse Compton scattering \citep{Lefa2012}, though in practice, 
this may be challenging to achieve with current instrumentation \citep{Romoli2017}.
The type of diffusion process can also have an impact on the temporal behaviour of the 
source. Hence, studying the light curve considering acceleration initiated by different 
diffusion processes will offer additional insights into the underlying characteristics.

\begin{acknowledgments}
This research has made use of data obtained from  NASA's High Energy Astrophysics Science Archive Research Center(HEASARC), a service of the Goddard Space Flight Center and the Smithsonian Astrophysical Observatory. CB wishes to thank CSIR, New Delhi (09/043(0198)/2018-EMR-I) for financial support. CB is thankful to UGC-SAP and FIST 2 (SR/FIST/PS1-159/2010) (DST, Government of India) for the research facilities in
the Department of Physics, University of Calicut. FMR acknowledges support by a DFG 
grant under RI 1187/8-1.
\end{acknowledgments}




\bibliographystyle{apsrev4-2}
\bibliography{ref} 



\label{lastpage}

\appendix

\begin{figure*}
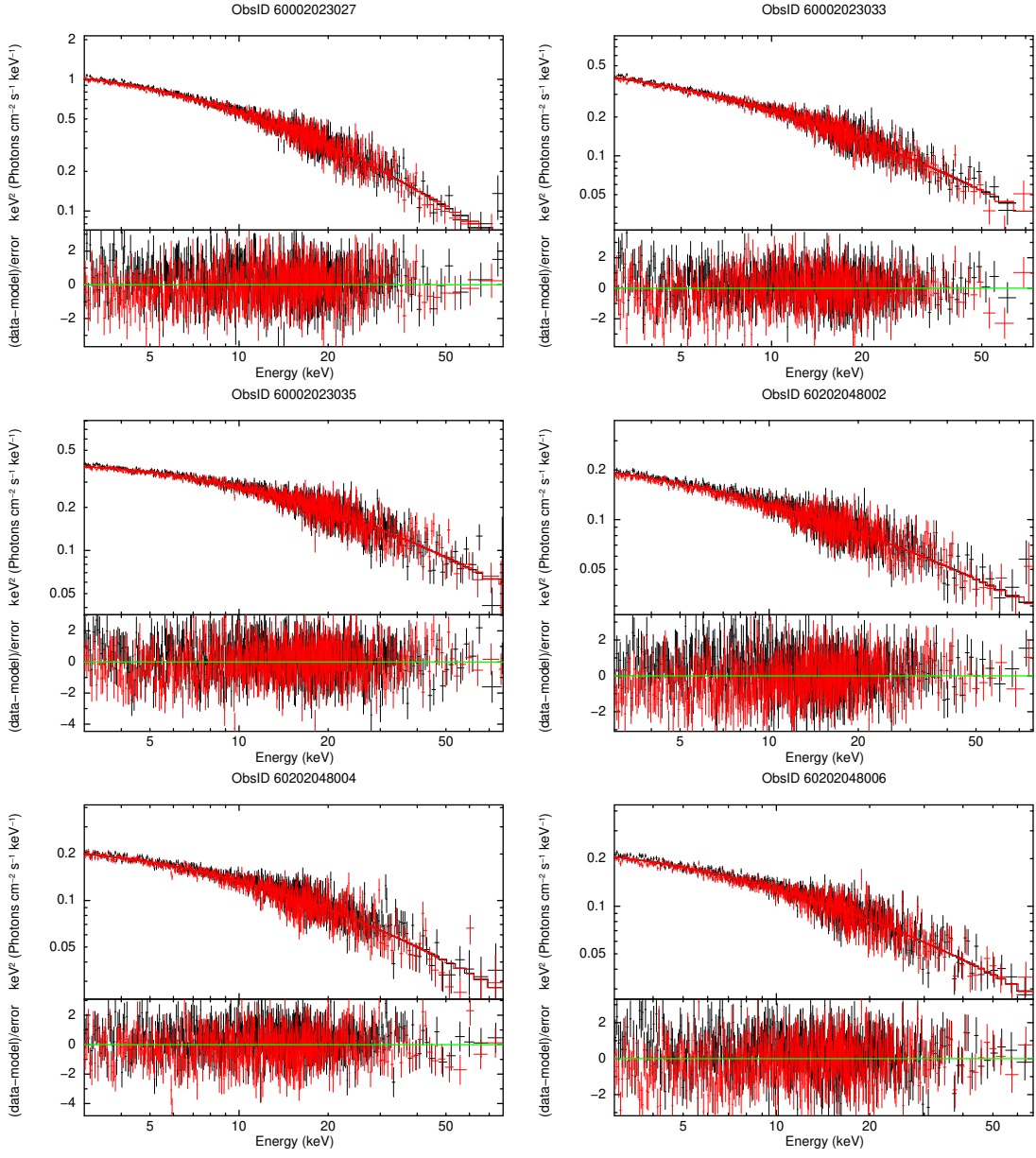

\begin{center}
\includegraphics[width=0.3\linewidth, angle =270]{nu-lp-60002023027-19-id.eps}
\includegraphics[width=0.3\linewidth, angle =270]{nu-lp-60002023033-22-id.eps}
\includegraphics[width=0.3\linewidth, angle =270]{nu-lp-60002023035-23-id.eps}
\includegraphics[width=0.3\linewidth, angle =270]{nu-lp-60202048002-26-id.eps}
\includegraphics[width=0.3\linewidth, angle =270]{nu-lp-60202048004-27-id.eps}
\includegraphics[width=0.3\linewidth, angle =270]{nu-lp-60202048006-28-id.eps}
\caption {\emph{NuSTAR} X-ray spectra (3-79 keV) along with log-parabola model.}
\label{fig:nu-logpar-fit-spec}  
\end{center}
\end{figure*}

\begin{figure*}
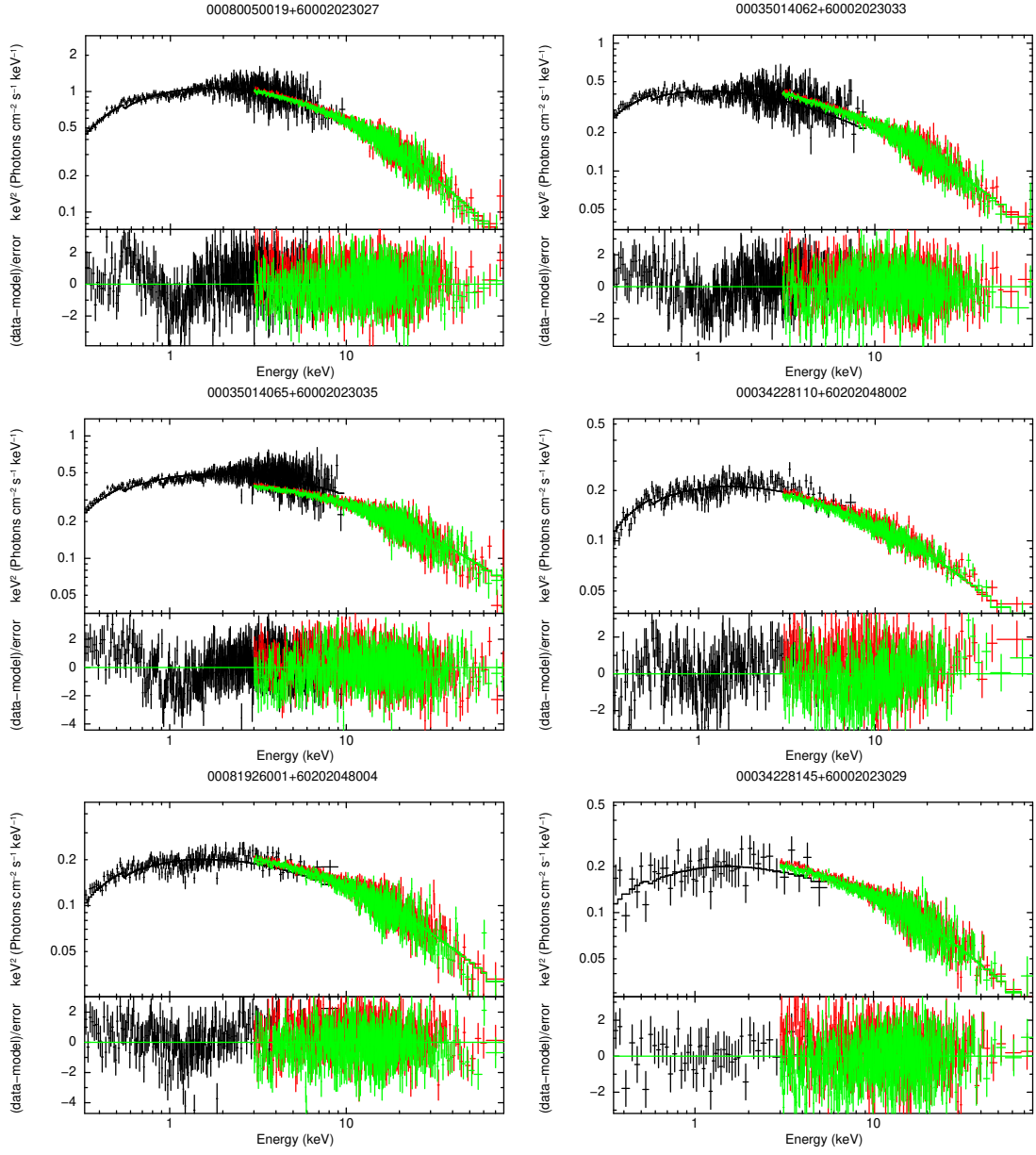

\begin{center}
\includegraphics[width=0.3\linewidth, angle =270]{19-lp-comb-label.eps}
\includegraphics[width=0.3\linewidth, angle =270]{22-lp-comb-label.eps}
\includegraphics[width=0.3\linewidth, angle =270]{23-lp-comb-label.eps}
\includegraphics[width=0.3\linewidth, angle =270]{26-lp-comb-label.eps}
\includegraphics[width=0.3\linewidth, angle =270]{27-lp-comb-label.eps}
\includegraphics[width=0.3\linewidth, angle =270]{28-lp-comb-label.eps}
\caption {Combined \emph{Swift}--XRT and \emph{NuSTAR}  X-ray spectra (0.3-79 keV) along with log-parabola model.}
\label{fig:comb-logpar-fit-spec}  
\end{center}
\end{figure*}

 \section{Sample X-ray spectral fits using a log-parabola model (Figure \ref{fig:nu-logpar-fit-spec}, \ref{fig:comb-logpar-fit-spec} ).}

\end{document}